\pdfoutput=1
\documentclass[11pt]{article}

\usepackage[final]{emnlp2021}

\usepackage{times}
\usepackage{latexsym}
\usepackage[T1]{fontenc}
\usepackage[utf8]{inputenc}

\usepackage{microtype}

\usepackage{times}  % DO NOT CHANGE THIS
\usepackage{helvet} % DO NOT CHANGE THIS
\usepackage{courier}  % DO NOT CHANGE THIS
\usepackage{graphicx} % DO NOT CHANGE THIS
\usepackage{natbib}  % DO NOT CHANGE THIS AND DO NOT ADD ANY OPTIONS TO IT
\usepackage{spreadtab}
\usepackage{diagbox}
\usepackage{caption} % DO NOT CHANGE THIS AND DO NOT ADD ANY OPTIONS TO IT
\usepackage[utf8]{inputenc}
\usepackage[T1]{fontenc}
\usepackage[english]{babel}
\usepackage{amsfonts}
\usepackage{nicefrac}
\usepackage{microtype}
\usepackage{amsmath}
\usepackage{amssymb}
\usepackage{mathtools}
\usepackage{subcaption}
\usepackage{multirow}
\usepackage{booktabs}
\usepackage{ragged2e}
\usepackage{tikz}
\usepackage{stackengine}
\usepackage{etoolbox}
\usepackage{xspace}
\usepackage{xpatch}
\usepackage{cuted}
\usepackage{enumerate}
\usepackage{xstring}
\usepackage{setspace}
\usepackage{tabularx}
\usepackage{makecell}
\usepackage{changepage}
\usepackage{enumitem}
\usepackage{cuted}
\usepackage{titlesec}
\usepackage{cancel}
\usepackage{eqparbox}
\usepackage{bibentry}
\usepackage[hang,flushmargin]{footmisc}
\usepackage[capitalise,noabbrev,nameinlink]{cleveref}
\usepackage[useregional=numeric]{datetime2}
\usepackage[linesnumbered,ruled,noend]{algorithm2e}
\usepackage{adjustbox}

\usepackage[switch,mathlines]{lineno}
\usepackage{lipsum}% http://ctan.org/pkg/lipsum
\usepackage{etoolbox}
\usepackage{longtable}
\graphicspath{{figures/}}
\title{Encoder Adaptation of Dense Passage Retrieval\\ for Open-Domain Question Answering}

\author{Minghan Li \and Jimmy Lin \\[1ex]
  David R. Cheriton School of Computer Science\\
  University of Waterloo\\[1ex]
  \texttt{\{m692li, jimmylin\}@uwaterloo.ca}}

\begin{document}
\maketitle
\begin{abstract}
One key feature of dense passage retrievers (DPR) is the use of separate question and passage encoder in a bi-encoder design.
Previous work on generalization of DPR mainly focus on testing both encoders in tandem on out-of-distribution (OOD) question-answering (QA) tasks, which is also known as domain adaptation.
However, it is still unknown how DPR's individual question/passage encoder affects generalization.
Specifically, in this paper, we want to know how an in-distribution (IND) question/passage encoder would generalize if paired with an OOD passage/question encoder from another domain.
We refer to this challenge as \textit{encoder adaptation}.
To answer this question, we inspect different combinations of DPR's question and passage encoder learned from five benchmark QA datasets on both in-domain and out-of-domain questions.
We find that the passage encoder has more influence on the lower bound of generalization while the question encoder seems to affect the upper bound in general. 
For example, applying an OOD passage encoder usually hurts the retrieval accuracy while an OOD question encoder sometimes even improves the accuracy.
\end{abstract}

\section{Introduction}
Generalization of neural networks has been a hot topic in various applications, such as computer vision~\citep{recht19imagenet,hendrycks19benchmarking} and natural language processing~\citep{talmor19multiqa,elsahar19annotate}.
In this paper, we are particularly interested in the generalization of dense passage retrievers (DPR)~\citep{karpukhin-etal-2020-dpr} in open-domain question answering~\citep{voorhees-tice-2000-trec,chen-etal-2017-reading}.
DPR leverages a bi-encoder structure which uses separate encoders for questions and passages and take the dot product of their vector output as relevance scores for end-to-end retrieval.

Previous work have been focusing on testing the generalization of DPR as a whole on out-of-distribution (OOD) data, which is also known as \textit{domain adaptation}~\cite{ramponi2021adaptation,nandan21beir}.
This line of work analyzes how DPR responds to different types of data, such as compositional questions~\citep{liu2021challenges}, novel entities~\cite{sciavolino2021entity}, and different languages~\citep{zhang21tydi}.
The common conclusion is that DPR transfers poorly to novel data distributions in zero-shot, compared to traditional term-matching algorithms such as BM25~\citep{robertson2009bm25}.
However, it remains unclear how individual question/passage encoder of DPR affects generalization in those tests.

\begin{figure}[t!]
\centering
\includegraphics[width=.49\textwidth]{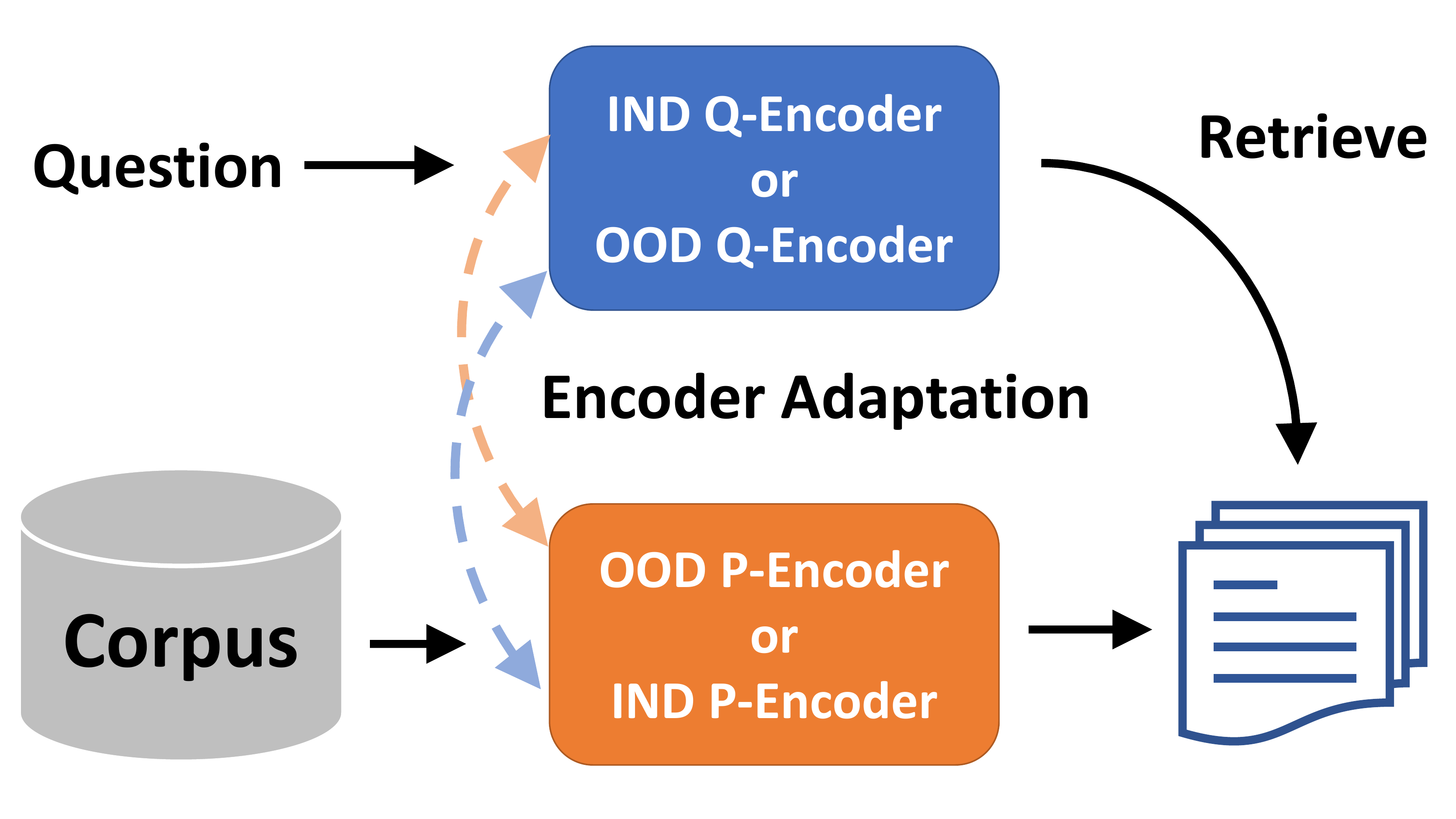}
\caption{The bi-encoder design of DPR enables one to individually test the generalization of the question (Q)/passage (P) encoder by plugging it into another DPR from a different domain, which we refer to as \textit{encoder adaptation}. ``IND'' means in-distribution and ``OOD'' means out-of-distribution.}
\label{fig:demo}
\end{figure}

In this paper, we aim to answer more fine-grained questions about generalization w.r.t. DPR: (1) Can a single in-distribution (IND) question/passage encoder generalizes well if paired with another OOD passage/question encoder from a different domain? (2) Do the question encoder and passage encoder play different roles in the generalization of DPR?
To clarify, an encoder is IND when tested on the domain where it is trained; an encoder is OOD when paired with another encoder and tested on a different domain.

To answer those questions, we evaluate various combinations of DPR's question encoder and passage encoder trained on five different question answering tasks (but using the same corpus, e.g., Wikipedia), which we refer to as ``encoder adaptation'' as shown in Fig.~\ref{fig:demo}.

For question (1), most IND $\times$ OOD pairs of question encoder and passage encoder under-perform the IND encoder pairs on most datasets.
However, there are some ``outlier'' encoder pairs that perform quite well on across all datasets. 
For example, the question encoder trained on CuratedTREC~\citep{baudivs2015curated} adapts reasonably well with different passage encoders, and the passage encoders trained on TriviaQA~\citep{joshi-etal-2017-triviaqa} and Natural Questions~\citep{kwiatkowski-etal-2019-nq} achieve acceptable accuracy when paired with different question encoders.

For question (2), we observe that the passage encoder seem to have a bigger impact on the generalization lower bound, while the question encoder seems to affect the upper bound more.
Specifically, if an IND question encoder is paired with an OOD passage encoder, then the performance would usually drop significantly on that dataset.
However, in the reverse setting (i.e., IND passage encoder $\times$ OOD question encoder), the performance does not suffer too much and sometimes even get improved.

For future work, such observations indicate that a pre-trained passage encoder can be fixed during adaptation and one could just fine-tune or re-train the question encoder to transfer DPR on other domains.
It is similar to the work from~\citet{sciavolino2021entity} and~\citet{zhan2021star} but for more general transfer learning, saving both time and space by avoiding repeatedly encoding dense indexes.

\section{Related Work}
\paragraph{Retrieval in QA} 
Term-matching methods such as tf-idf or BM25~\citep{robertson2009bm25,lin2021pyserini} has established strong baseline in various QA tasks~\citep{chen-etal-2017-reading,yang-etal-2019-end-end,min-etal-2019-discrete}. 
Recently, retrievers based on neural networks~\citep{goodfellow16deep} and pre-trained language models~\citep{devlin-etal-2019-bert} also make great advancement in open-domain question-answering~\citep{seo-etal-2019-real,lee-etal-2019-latent,guu2020realm}.
Particularly, dense passage retrieval (DPR)~\citep{karpukhin-etal-2020-dpr} sets the milestone by encoding questions and passages separately with a bi-encoder design.
Based on DPR, multiple work on compression~\citep{yamada-etal-2021-bpr,Izacard20efficientqa}, hard-negative mining~\citep{xiong21ance,zhan2021star}, and QA pre-training~\citep{lu21less,gao21condenser} has further pushed the performance boundary of in-distribution dense retrieval.

\paragraph{Domain Adaptation in Retrieval} BEIR investigates DPR's transferability over multiple retrieval tasks containing different knowledge~\citep{nandan21beir}, while Mr.TYDI evaluates DPR pretrained on English corpus in a multi-lingual setting~\citep{zhang21tydi}. 
The zero-shot cross-domain/language results of DPR are unsatisfactory as expected, while other works on in-distribution generalization also find that DPR performs poorly on certain types of questions.
\citet{liu2021challenges} observes that neural-retrievers fail to generalize to compositional questions and novel entities.
\citet{sciavolino2021entity} also finds that dense models can only generalize to common entities or certain question patterns.

\begin{table*}[t]
\centering
\resizebox{15.5cm}{!}{
  \begin{tabular}{l@{\hskip 0.1in}|c@{\hskip .2in}c@{\hskip .2in}c@{\hskip .2in}c@{\hskip .2in}c@{\hskip 0.2in}}
\toprule

% \textbf{Question-Index Pair} 
\diagbox[width=7.5em]{\textbf{Encoder}}{\textbf{Test set}} &NQ-test &Trivia-test &WQ-test &Curated-test &SQuAD-test \\
\midrule
BM25-Q-P &62.9/78.3  &76.4/83.2 &62.4/75.5 &80.7/89.9 &\textbf{71.1}/\textbf{81.8} \\
\midrule
NQ-Q-P &\textbf{79.1}/\textbf{85.9} &69.1/78.6 &67.4/78.3 &85.3/91.4 &47.7/63.3 \\
Trivia-Q-P &62.5/72.4 &\textbf{78.9}/\textbf{84.5} &70.0/\textbf{80.6} &\textbf{88.2}/\textbf{92.9} &53.2/68.7 \\
WQ-Q-P &46.3/61.9 &58.0/71.4 &\textbf{71.0}/80.2 &75.8/86.5 &41.4/59.1 \\
Curated-Q-P &56.0/68.0 &69.0/79.7 &66.8/77.9 &85.1/92.2 &50.4/66.1 \\
SQuAD-Q-P &41.4/61.6 &63.7/77.3 &57.9/74.1 &79.5/91.1 &62.1/76.8 \\
\bottomrule
  \end{tabular}
  }
  \caption{IND/OOD Q-P encoder pairs: Top-20/Top-100 retrieval accuracy (\%) on five benchmark QA test sets.
  Each score represents the percentage of top 20/100 retrieved passages that contain answers. 
  ``NQ-test'' means the test data of NQ; ``NQ-Q-P'' means using the DPR's question and passage encoder trained on NQ to encode the test questions and Wikipedia corpus.
}
    \label{tbl:results_zeroshot}
\end{table*}

\section{Dense Passage Retrieval}\label{section:background}

\paragraph{Retrieval/Inference} Given a corpus of passages $\{p_1, p_2, \cdots, p_n\}$ and a query $q$, DPR~\citep{karpukhin-etal-2020-dpr} leverages two encoders $f_Q$ and $f_P$  to encode the question and documents, respectively.
The similarity between the question $q$ and document $p$ is defined as the vector dot product: \begin{align}\label{eq:dpr_sim}
    s = E_q^TE_p,
\end{align}
where $E_q$ and $E_p$ are the output of $f_Q$ and $f_P$, respectively.
The similarity score $s$ will be used to rank the passages during retrieval.
Both $f_Q$ and $f_P$ use the pre-trained BERT~\citep{devlin-etal-2019-bert} for initialization and use the \texttt{[CLS]} vector as the representation.

\paragraph{Training} As pointed out by~\citet{karpukhin-etal-2020-dpr}, training the encoders such that EQ~\eqref{eq:dpr_sim} becomes a good ranking function is essentially a metric learning problem~\citep{kulis2012metric}.
Given a specific question $q$, let $p^+$ be the positive context that contains answers for $q$ and $P=\{p_1^-, p_2^-,...p_k^-\}$ be the negative contexts, 
the negative log likelihood objective w.r.t. $q$ and $P$ is:
\begin{align}\label{eq:dpr_obj}
\centering
    &\mathcal{L}(q, p^+,p_1^-, p_2^-,...p_k^- )\nonumber\\
    =&-\log \frac{\exp(E_q^TE_{p^+})}{\exp(E_q^TE_{p^+}) + \sum\limits_{i=1}^{k} \exp(E_q^TE_{p_i^-})}.
\end{align}

\section{Experimental Setup}\label{section:setup}
We follow the DPR paper~\citep{karpukhin-etal-2020-dpr} to train and evaluate our dense retrievers. 
We replicate their results on 5 benchmark datasets, with a maximum score difference between ours and their numbers of 1\%.
This work only focuses on evaluating the retrieval performance and therefore we do not include the reader module into consideration. 
We report the top-20 and top-100 recall accuracy as the metrics for evaluation.

\paragraph{Datasets} We train individual DPR models on 5 standard benchmark QA tasks: Natural Questions (NQ)~\citep{kwiatkowski-etal-2019-nq}, TriviaQA (Trivia)~\citep{joshi-etal-2017-triviaqa}, WebQuestions (WQ)~\citep{berant-etal-2013-wq}, CuratedTREC (TREC)~\citep{baudivs2015curated}, SQuAD-1.1 (SQuAD)~\citep{rajpurkar2016squad} as shown in Tbl.~\ref{tbl:data_size}.
We evaluate the retriever models on the test sets of the aforementioned datasets. 

For retrieval, we chunk the Wikipedia collections~\citep{guu2020realm} into passages of 100 words as in~\citet{wang2019multi}, which yields about 21 million samples in total.
We follow~\citet{karpukhin-etal-2020-dpr} using BM25~\citep{robertson2009bm25,lin2021pyserini} to select the positive and negative passages.

\begin{table}[t]
\centering
\resizebox{7.5cm}{!}{
  \begin{tabular}{l@{\hskip 0.1in}r@{\hskip .2in}r@{\hskip .2in}r@{\hskip .2in}}
\toprule

% \textbf{Question-Index Pair} 
\textbf{Datasets} &Train &Dev &Test\\
\midrule
Natural Questions  &58,880 &8,757 &3,610\\
TriviaQA  &60,413 &8,837 &11,313\\
WebQuestions  &2,474 &361 &2,032\\
CuratedTREC  &1,125 &133 &694\\
SQuAD  &70,096 &8,886 &10,570\\
\bottomrule
  \end{tabular}
  }
  \caption{Number of questions in each QA dataset from~\citet{karpukhin-etal-2020-dpr}. The
column of Train denotes the number questions after filtering.
}
    \label{tbl:data_size}
\end{table}

\paragraph{Models and Training}\label{section:experiments:model}
We train individual DPR models on the training set of NQ, TriviaQA, WQ, CuratedTREC, and SQuAD-1.1 separately following~\citet{karpukhin-etal-2020-dpr}. 
We optimize the objective function in EQ~\eqref{eq:dpr_obj} with learning rate of 2e-05 using Adam~\citep{DBLP:journals/corr/KingmaB14} for 40 epochs.
The rest of the hyperparameter setting remains the same as described in~\citet{karpukhin-etal-2020-dpr}.

\begin{table*}[t]
\centering
\resizebox{15.5cm}{!}{
  \begin{tabular}{l@{\hskip 0.1in}|c@{\hskip 0.1in}c@{\hskip 0.1in}c@{\hskip 0.1in}c@{\hskip 0.1in}c@{\hskip 0.1in}}
\toprule

\diagbox[width=8.5em]{\textbf{P-Encoder}}{\textbf{Q-Encoder + }\\\textbf{Test set}}&NQ-test-Q &Trivia-test-Q &WQ-test-Q &Curated-test-Q &SQuAD-test-Q \\
\midrule
NQ-P-encoder &\textbf{79.1}/\textbf{85.9} &71.6/80.6 &66.9/77.5 &87.5/92.5 &51.2/67.3\\
Trivia-P-encoder &68.8/79.7 &\textbf{78.9}/\textbf{84.5} &65.7/77.7&\textbf{88.8/94.2}&53.0/69.1\\
WQ-P-encoder &55.1/69.3 &66.5/77.2 &\textbf{71.0}/\textbf{80.2} &81.8/91.4&49.3/65.7\\
Curated-P-encoder &58.5/71.9&67.3/77.9&61.9/74.4&85.1/92.2 &48.8/66.2 \\
SQuAD-P-encoder &54.6/70.9&67.3/78.9&55.7/71.8&81.5/92.3 &\textbf{62.1}/\textbf{76.8} \\
\bottomrule
  \end{tabular}
  }
  \caption{IND Q-encoder $\times$ OOD P-encoder: Top-20/Top-100 retrieval accuracy (\%) on benchmark QA test sets.
  ``XX-test-Q'' means we use the ``XX'' test set and the question encoder trained on the ``XX'' training set. ``YY-P-encoder'' means we use the passage encoder trained from ``YY'' training set.
}
    \label{tbl:results_question}
\end{table*}

\begin{table*}[t]
\centering
\resizebox{15.5cm}{!}{
  \begin{tabular}{l@{\hskip 0.1in}|c@{\hskip 0.1in}c@{\hskip 0.1in}c@{\hskip 0.1in}c@{\hskip 0.1in}c@{\hskip 0.1in}}
\toprule

\diagbox[width=8.5em]{\textbf{Q-Encoder}}{\textbf{P-Encoder + }\\\textbf{Test set}}&NQ-test-P &Trivia-test-P &WQ-test-P &Curated-test-P &SQuAD-test-P \\
\midrule
NQ-Q-encoder &79.1/85.9 &73.9/82.0 &70.2/80.1&83.3/90.3&55.1/72.3\\
Trivia-Q-encoder &75.9/84.8&78.9/84.5&72.0/81.7&82.6/92.1&59.3/74.9\\
WQ-Q-encoder &70.3/81.6&70.6/80.1&71.0/80.2&79.3/89.6&54.3/71.1\\
Curated-Q-encoder &\textbf{79.2/86.1} &\textbf{78.9}/\textbf{84.8} &\textbf{74.1}/\textbf{82.1}&\textbf{85.1}/\textbf{92.2}&59.7/75.4\\
SQuAD-Q-encoder &73.7/83.9 &75.5/83.4 &71.7/81.2&82.4/92.4&\textbf{62.1}/\textbf{76.8} \\
\bottomrule
  \end{tabular}
  }
  \caption{IND P-encoder $\times$ OOD Q-encoder: Top-20/Top-100 retrieval accuracy (\%) on benchmark QA test sets.
  ``XX-test-P'' means we use the ``XX'' test set and the passage encoder trained on the ``XX'' training set. ``YY-Q-encoder'' means we use the question encoder trained from ``YY'' training set.
}
    \label{tbl:results_index}
\end{table*}

\section{Can IND Q/P-Encoders Generalize to OOD P/Q-Encoders?}
\subsection{IND/OOD Q-P Encoder Pairs}
Tbl.~\ref{tbl:results_zeroshot} shows the zero-shot retrieval performance using different DPR models and BM25 on 5 benchmark QA datasets.
Each question encoder is paired with its original passage encoder and evaluated on different test sets.
Normally, the in-distribution DPR model is expected to outperform the OOD DPR, which is the situation that happens to most datasets such as NQ, Trivia, and SQuAD.
However, for datasets such as WQ and Curated, we find that the DPR trained on Trivia has better zero-shot performance than the IND ones.
We suspect the reason is that NQ and Trivia has much larger training data than WQ and Curated as shown in Tbl.~\ref{tbl:data_size}, which potentially includes some similar questions in Curated and WQ
Moreover, BM25 outperforms the DPR model on SQuAD as SQuAD mainly contains entity-centred questions which is good for term-matching algorithms.
In constrast, DPR has been shown to have poor generalization on novel entities~\citep{liu2021challenges,sciavolino2021entity}, and therefore the OOD DPR perform poorly on SQuAD in general.

\subsection{IND Q-Encoder $\times$ OOD P-Encoder}\label{sec:results:idq_oodp}
Tbl.~\ref{tbl:results_question} shows the adaptation performance using IND question encoders and OOD passage encoders trained on other domains.
The performance immediately drops once the IND question encoder is paired with an OOD passage encoders on most datasets.
However, compared to the zero-shot setting in Tbl.~\ref{tbl:results_zeroshot}, most OOD passage encoders have significant improvement when paired with the IND question encoders, except for WQ where all OOD passage encoders have significant accuracy drop when paired with the WQ question encoder.
This indicates that the representations learned from the WQ data are shifted towards some weird manifolds that do no align well with other encoders.
In contrast, the passage encoders trained on NQ and Trivia still perform well on most datasets, which supports the argument that NQ and Trivia learns more general representations due to learning from more diverse training data.

\subsection{IND P-Encoder $\times$ OOD Q-Encoder}
Tbl.~\ref{tbl:results_index} shows the adaptation performance using IND passage encoders and OOD question encoders trained on other domains.
Similar to the previous section, the IND passage encoders significantly improve the generalization compared to the zero-shot setting.
The difference to Section~\ref{sec:results:idq_oodp} is that the accuracy is more robust against different questions encoders with the IND passage encoder.

Surprisingly, when the IND passage encoders are paired with particular OOD question encoders, the accuracy is sometimes even better than the in-distribution counterpart!
For example, the Curated question encoder outperforms the IND question encoder on NQ, Trivia, WQ, and Curated when paired with the in-distribution passage encoder of each domain.
Therefore, we propose two conjectures from the results in Tbl.~\ref{tbl:results_question} and Tbl.~\ref{tbl:results_index}: 
(1) Using in-domain passage encoders is more important than using in-domain question encoders. 
(2) Curated question encoder learns more general question representations that align well with others with much less training data.
We conjecture that it is because Curated covers more question patterns or entities that might help to learn more general question encoders.

\section{Does Q/P-Encoder Play a Different Role in the Generalization of DPR?}
\begin{figure}[t!]
\centering
\includegraphics[width=.48\textwidth]{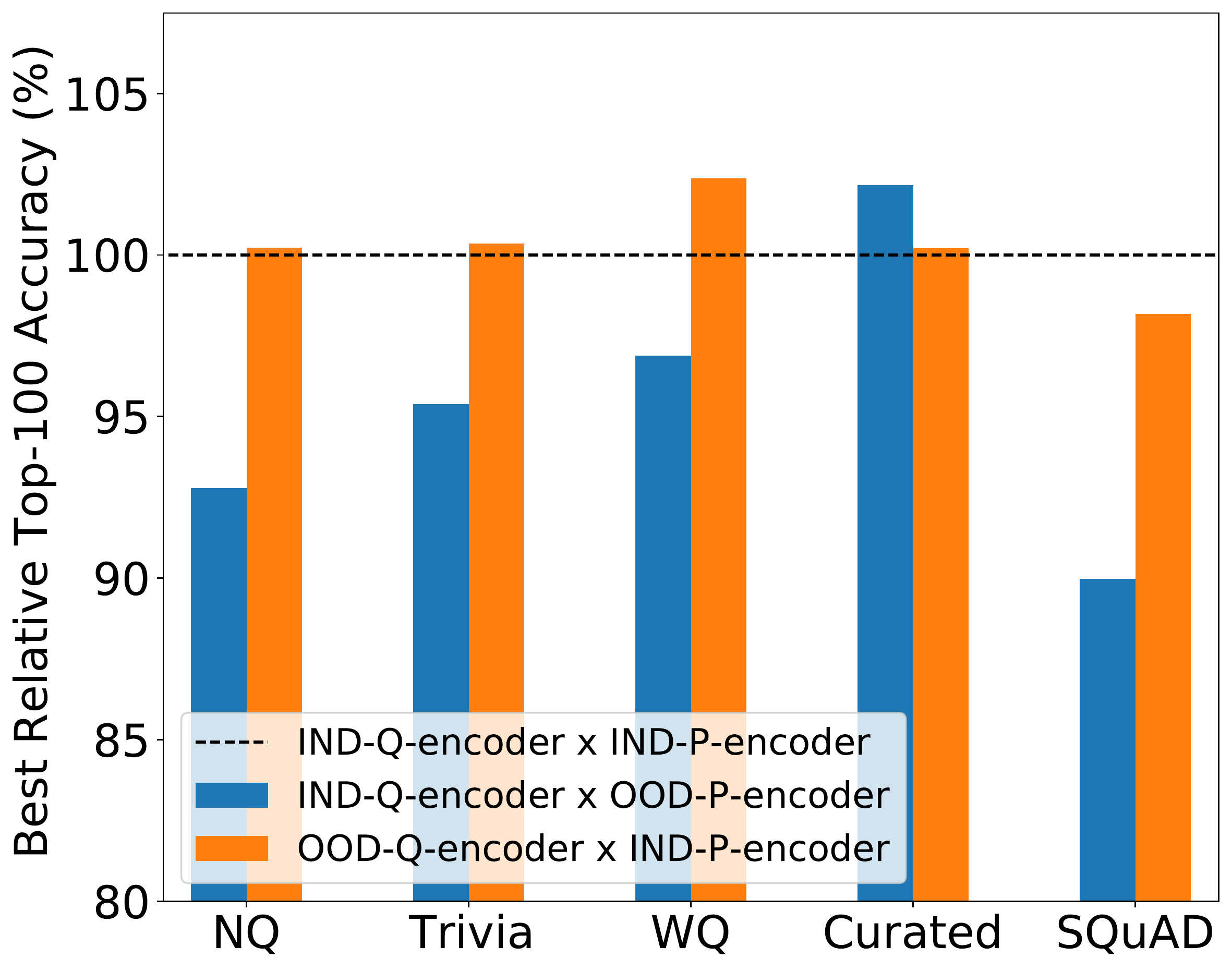}
\caption{Comparison between IND Q-encoder $\times$ OOD P-encoder and OOD Q-encoder $\times$ IND P-encoder. The x-axis represents different test sets, and the y-axis represents the best top-100 accuracy (\%) relative to the IND question-passage encoder pair by replacing either the question or passage encoder with another one.}
\label{fig:lower_upper}
\end{figure}

To analyze the role that the question encoder and passage encoder each plays in the generalization of DPR, we compare the relative top-100 retrieval test accuracy between using the best OOD question encoder and the best OOD passage encoder for each dataset as shown in Fig.~\ref{fig:lower_upper}. 
We can see that using an OOD passage encoder consistently underperforms using an OOD question encoder (except for Curated which has the best OOD question encoder as well, making the passage encoder bar looks like an outlier).
It suggests that the passage encoder is the key to DPR's generalization as it lower-bounds the IND question-passage encoder pair's performance most of the time.

From the last section, using an OOD question encoder does not hurt the retrieval accuracy too much but sometimes even outperforms using an IND question encoder.
Fig.~\ref{fig:lower_upper} shows that using a best OOD question encoder outperforms the IND question-passage encoder pairs on most dataset, except for SQuAD whose data are biased towards certain data types (e.g., named entities). 
Such results are consistent with the findings in hard-negative mining~\citep{zhan2021star} and entity transfer learning~\citep{sciavolino2021entity} for DPR.

\section{Conclusions}
We address two questions regarding the generalization of DPR: (1) Can its question/passage encoder generalizes to another passage/question encoder of another DPR trained on a different domain? (2) What roles do the question encoder and passage encoder play in the generalization of DPR?  
To answer these questions, we examine different combinations of DPR's question encoders and passage encoders trained on five different QA datasets, which we refer to as ``encoder adaptation''.
Despite most pairs of question encoder and passage encoder from different sources underperform the in-distribution DPR model, we find the passage encoders learned on Natural Questions and TriviaQA achieve reasonable accuracy due to more training data.
In addition, the question encoder from the CuratedTREC dataset adapts particular well with different passage encoders, whose performance even outperform the in-distribution DPR model.
Finally, we observe that the passage encoder affects the generalization lower bound more while the question encoder seems to play a more vital role in the generalization upper bound, indicating the possibility of transferring DPR to another domain by just fine-tuning the question encoder while keeping the passage encoder fixed.

\section*{Acknowledgements}

This research was supported in part by the Canada First Research Excellence Fund and the Natural Sciences and Engineering Research Council (NSERC) of Canada; computational resources were provided by Compute Ontario and Compute Canada.

\bibliography{anthology,custom}
\bibliographystyle{acl_natbib}

\end{document}